\documentclass[10pt,showpacs,superscriptaddress,twocolumn,aps, longbibliography]{revtex4-1}

\usepackage{hyperref}
\usepackage{graphicx}
\usepackage{dcolumn}
\usepackage{bm}
\usepackage[utf8]{inputenc}
\usepackage{transparent}
\usepackage[detect-family = true,detect-inline-family = math, detect-weight = true,detect-inline-weight = math,separate-uncertainty=true, multi-part-units = single, product-units=single]{siunitx}
\usepackage{xcolor}
\usepackage{float}
\usepackage{xfrac}
\DeclareSIUnit\year{yr}
\graphicspath{{figures/}}
\usepackage{xr-hyper}
\usepackage{upgreek}

\renewcommand{\vec}[1]{\boldsymbol{#1}}

\begin{document}
\title{Nano-structured alkali-metal vapor cells}

\author{T. F. Cutler*}
\affiliation{Joint Quantum Centre (JQC) Durham-Newcastle, Department of Physics, Rochester Building, Durham
University, South Road, Durham, DH1 3LE}
\author{W. J. Hamlyn*}
\affiliation{Joint Quantum Centre (JQC) Durham-Newcastle, Department of Physics, Rochester Building, Durham
University, South Road, Durham, DH1 3LE}
\author{J. Renger}
\affiliation{Max Planck Institute for the Science of Light, 91058 Erlangen, Germany}
\author{K. A. Whittaker}
\affiliation{Joint Quantum Centre (JQC) Durham-Newcastle, Department of Physics, Rochester Building, Durham
University, South Road, Durham, DH1 3LE}
\author{D. Pizzey}
\affiliation{Joint Quantum Centre (JQC) Durham-Newcastle, Department of Physics, Rochester Building, Durham
University, South Road, Durham, DH1 3LE}
\author{I. G. Hughes}
\affiliation{Joint Quantum Centre (JQC) Durham-Newcastle, Department of Physics, Rochester Building, Durham
University, South Road, Durham, DH1 3LE}
\author{V. Sandoghdar}
\affiliation{Max Planck Institute for the Science of Light, 91058 Erlangen, Germany}
\affiliation{Friedrich Alexander University Erlangen-Nuremberg, D-91058 Erlangen, Germany}
\author{C. S. Adams}
\affiliation{Joint Quantum Centre (JQC) Durham-Newcastle, Department of Physics, Rochester Building, Durham
University, South Road, Durham, DH1 3LE}

\date{\today}

\begin{abstract}
Atom-light interactions in nano-scale systems hold great promise for novel technologies based on integrated emitters and optical modes. We present the design architecture, construction method, and characterization of an all-glass alkali-metal vapor cell with nanometer-scale internal structure. Our cell has a glue-free design which allows versatile optical access, in particular with high numerical aperture optics, and incorporates a compact integrated heating system in the form of an external deposited ITO layer. By performing spectroscopy in different illumination and detection schemes, we investigate atomic densities and velocity distributions in various nanoscopic landscapes. We apply a two-photon excitation scheme to atoms confined in one dimension within our cells, achieving resonance line-widths more than an order of magnitude smaller than the Doppler width. We also demonstrate sub-Doppler line-widths for atoms confined in two dimensions to micron-sized channels. Furthermore, we illustrate control over vapor density within our cells through nano-scale confinement alone, which could offer a scalable route towards room-temperature devices with single atoms within an interaction volume. Our design offers a robust platform for miniaturized devices that could easily be combined with integrated photonic circuits. 
\end{abstract}

\maketitle

\section{Introduction}

Individual atoms or atomic ensembles offer an attractive platform for quantum sensing and devices. Atom-light interactions have already found use in applications such as clocks~\cite{Knappe2005,Boyd2007}, optical filtering~\cite{Zentile2015,Elecsus2018,Keaveney2018}, magnetometry~\cite{Bison2003,Schwindt2004,Balabas2006,Budker2007,Bison2009,Wyllie2012,Shah2013,Sheng2013,Boto2017,Klinger2019}, and laser stabilization~\cite{Keaveney2016,Mathew2018}. However, most atomic devices are either macroscopic or at best mesoscopic~\cite{Budker2007}. For some applications, it would be attractive to achieve nano-scale confinement or localization, which is standard for solid-state systems~\cite{Maze2008} but challenging for atoms.

Indeed, work towards confinement of atoms and light fields on the micro- and nano-scales has increasingly offered diverse access to studying fundamental physics~\cite{Sandoghdar1992,Sukenik1993,Zambon1997,Keaveney2012,Petersen2014,LeKien2015,Lodahl2017,LeKien2017,Peyrot2018}, sensing~\cite{Briaudeau1999,Bloch2005,Fichet2007,Bison2009,Kitching2011,Peyrot2019}, miniaturization of optical devices~\cite{Schwindt2004,Knappe2005,Sander2012}, and quantum technologies~\cite{Lounis2005,Jiang2007,Kimble2008,Lvovsky2009,Ripka2018,Wang2019}. Although the first atom-on-chip attempts date back about two decades~\cite{Reichel_1999,Kiel_2016_review}, the difficulties of taming surface effects such as van der Waals forces~\cite{Hinds1991, Henkel1998, Whittaker2014} have rendered the control of cold atoms at nanoscopic distances elusive~\cite{Aoki2006}.

The continued study of atom-light interaction in thermal vapors has offered diverse access to fundamental physics~\cite{Marek1979,Walker2012,Whiting2018}, and miniaturization of such thermal vapor systems is an emerging alternative approach to atomic physics at the micrometer- and nanometer-scale. Examples include confinement of vapors to micron-sized hollow core fibers~\cite{Bhagwat2008} and construction of sub-micron alkali-metal vapor cells~\cite{Sarkisyan2001,Baluktsian2010,Whittaker2015,Peyrot2019}. Developments in miniaturized alkali-metal vapor cell technology on the millimeter scale have already enabled new atom-based devices, such as chip-scale atomic clocks~\cite{Liew2004} and magnetometers~\cite{Schwindt2004}. Promising further technological developments on the micrometer-scale include integrated diffractive elements~\cite{Liron2019}. As well as the fundamental interest in atom-light interactions at the nano-scale already highlighted, there is also interest in interfacing thermal vapors with nanophotonics~\cite{Stern2013,Stern2016,Ritter2016,Ritter_2018} and integrated opto-mechanical features~\cite{Midolo2018}. However, much work remains to be done before nanometer-scale vapor-based devices can be used in technological applications, e.g.\@ for local sensing~\cite{Klinger2019} or quantum networks~\cite{Kimble2008,Reiserer2015}. In particular, reliable fabrication methods are needed for the realization of scalable and efficient platforms. Furthermore, the interactions of the atoms with the cell walls~\cite{Hinds1991,Sukenik1993,Bloch2005,Fichet2007,Jentschura2016} as well as their velocity distribution and diffusion behavior have to be investigated and understood.

In this work we present a new platform that produces long lifetime (at least 1 year) alkali-metal vapor nano-cells. Using laser lithography and reactive ion etching we create regions with internal dimensions from $\SI{2}{\micro \metre}$ to $\SI{400}{\nano\metre}$ (though smaller structures with dimension below 100~nm are possible), filled with rubidium vapor. We demonstrate full optical access that facilitates the use of conventional high numerical aperture (NA) optics for efficient atom-light coupling. Previously reported nano-cells are generally wedge-shaped, with confinement in one spatial dimension decreasing continuously to \SI{0}{\micro\meter}~\cite{Sarkisyan2001,Peyrot2019}. This wedge design, in combination with thick cell windows, does not allow for such versatile optical access or use of high-NA optics, nor near-arbitrary patterned confinement geometries. In contrast, existing works demonstrate the use of hollow-core fibers for confinement of atomic vapors in not one but two spatial dimensions. However, these are limited to core sizes of order microns. In this study, we show that our nano-cell etching process presents a significant improvement, producing atom confinement in one and two dimensions with sub-micron lengthscales. For the reasons outlined above, our novel design and manufacturing process presents a significant advancement in nano-cell and confined-vapor technology. We illustrate the benefits and capabilities of our nano-cell by experimentally demonstrating two excitation and detection schemes for atoms confined within various nano-scale structures. Our approach opens doors to efficient and compact architectures for performing quantum nano-optical studies.

\section{Nano-cell Fabrication}
\begin{figure}
    \centering
    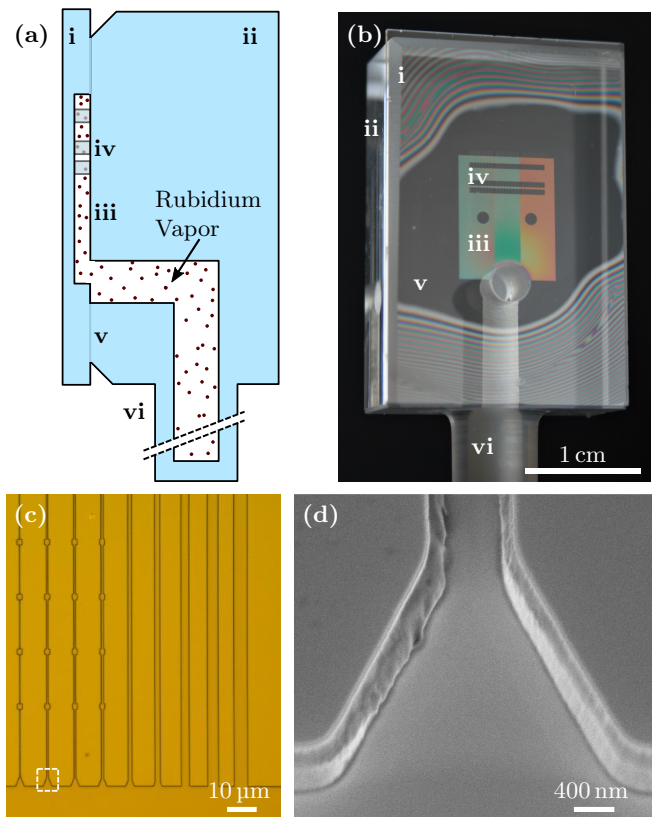
    \caption{\textbf{(a)} Side cross-section schematic (not to scale) and \textbf{(b)} front-facing photograph of the nano-cell. The etched cover slide (i) is contacted onto the block (ii) such that the etched patterns form a nano-thickness region (iii) and nano-channels (iv) between the two glass pieces. An optical contact bond (v) surrounds the etched area to act as a vacuum seal. The rubidium reservoir below (vi) is attached to provide atomic vapor into the nano-structured region. This cell contains areas with confinement in both one and two dimensions. \textbf{(c)}~Front-facing bright field microscopy images of various channel structures, with widths~$\SIrange{400}{5000}{\nano\metre}$. A number of copies of these structures are etched into the cover slide to provide a range of depths~($\SIrange{400}{2000}{\nano\metre}$). These channel structures reside in the horizontal black bars, labelled (iv) in (b). Vapor is supplied from below in this image. \textbf{(d)}~SEM image of a single channel aperture (indicated by the white box in (c)).
    }
    \label{fig:cell_photo}
\end{figure}

Figure~\ref{fig:cell_photo}(a) displays the schematics of our glass nano-cells, depicting a side view of the cell region, with nanoscopic features located at positions (iii) and (iv). A tube conduit (diameter 5\,mm, labelled (vi)) connects this nano-structured region to an alkali-metal reservoir. The glass cell is assembled from three distinct parts that we refer to as: the cover slide (labelled (i)); the block (labelled (ii)); and the reservoir (located below label (vi)). In this work, the block was a fused silica cuboid of dimensions $\SI{10}{\milli \metre} \times \SI{20}{\milli \metre} \times \SI{30}{\milli \metre}$, though this was chosen for manufacturing convenience and could be considerably reduced in size for future applications. The cover slide is a fused silica panel of dimensions $\SI{0.5}{\milli \metre} \times \SI{20}{\milli \metre} \times \SI{30}{\milli \metre}$ into which we ``write'' the nano-structures of choice. To do this, we pattern micro- and nano-scale features into photoresist using direct-write laser lithography, and transfer these to the glass substrate using reactive ion etching~\cite{Gmeiner:16}. Figure~\ref{fig:cell_photo}(b) shows a photograph of the cell under white-light illumination (full details of cell characterization will be discussed in Section~\ref{sec:char}).

The block is optically polished on all sides to increase optical access to the nanoscopic regions, and to facilitate the formation of an optical contact bond (OCB). We form an OCB between the cover slide and the block to act as a vacuum seal encircling the patterned area. The cover slide is placed on the block such that the micro- and nano-patterns form a confined region between the two pieces, which also overlaps the internal tube of the block to allow vapor to enter. Figure~\ref{fig:cell_photo}(b) shows the OCB surrounding the patterned area, seen as lack of Newton's fringes between the two fused silica pieces. Once an OCB encircling the nano-regions and milled aperture has been formed, the bond is made permanent by firing in a kiln at $\SI{1000}{\degreeCelsius}$ for $\SI{6}{\hour}$ at atmospheric pressure. The OCB process up until firing is reversible. Two fused silica spacer columns (labelled in Fig.\@~\ref{fig:WL_data}(a)) were included in our cell design to prevent the cavity thickness from deviating from the design specification due to bowing during the next fabrication step, when the air is pumped out to produce a vacuum.

To load the nano-cell with alkali-metal atoms (in our case rubidium at natural abundance) the fired pieces are attached to a borosilicate glass manifold by glassblowing. A vacuum pump and a rubidium break seal ampoule are attached to the manifold, which is evacuated and generally sealed off at $10^{-6}$~\SI{}{\milli\bar}. The rubidium ampoule is broken and, by melting with a low flame, liquid rubidium is made to flow towards the block. A short (typically~\SI{5}{\centi\meter}) section of the manifold immediately adjacent to the block, and containing condensed rubidium, is then sealed and detached to enclose the complete nano-cell. This section of borosilicate glass tube attached to the bottom of the block is referred to as the reservoir (see Fig.~\ref{fig:cell_photo}(a), part (vi)). Note that the cell is not baked during the evacuation process. However, spectroscopic measurements (such as those shown later in this study) do not show prohibitive resonance line-shifts or broadening, and thus we do not consider residual background gas to be a problem.

We find that our glue-free design is very robust, and we have produced numerous nano-cells with an operation lifetime of over $\SI{1}{\year}$. The cells can withstand dozens of heating and cooling cycles between room temperature and $\SI{180}{\degreeCelsius}$. In typical operation both the cell block and the reservoir are independently heated, so as to maintain a thermal gradient between the two and thus avoid rubidium condensation in the nano-structured regions. Heating is achieved via patches of sputtered indium tin oxide (ITO) on external surfaces of the cell, which act as resistive heaters when an electric current is applied (not shown in Fig.~\ref{fig:cell_photo}(b)). Experiments with thermal vapors often rely on bulky ovens for cell heating, but the compactness and transparency of our ITO layer allows heating of the cell without compromising optical access or the experimental footprint. 

To populate the nano-scale regions with atoms, we find that it is favourable to periodically `flood' these areas with liquid rubidium by inducing a reverse temperature gradient (reservoir hotter than the nano-region), and then returning to a conventional temperature gradient to evacuate the nano-regions again. After such a cycle deposits of rubidium remain on the inner surfaces which act as local reservoirs.

\section{Characterization} \label{sec:char}
 Figure~\ref{fig:cell_photo}(b) shows a completed nano-cell with the patterned regions made visible by white light interference in the void formed between the two fused silica pieces, producing strong colours. More subtle, however, is the fact that the colors change both gradually and in discrete steps across the patterned area of the nano-cell. The discrete color changes occur by design due to the discrete depth changes selected when patterning the cover slide. The gradual color changes are a result of the surface of the block and the internal surface of the cover slide being non-parallel. We also note that the gradients generally differ before and after the kiln firing process, suggesting the process induces changes in mechanical stress in the two fused silica pieces.

To characterize the cells, the thickness of the patterned regions is measured after the assembly is complete. To do so, we exploit the fact that the nano-thickness region acts as a low finesse Fabry-P\'erot cavity, with the internal surfaces of the fused silica cavity acting as low reflectivity mirrors. The transmission, $T$, of a Fabry-P\'erot interferometer is~\cite{Opticsf2f}
\begin{equation}
T = \frac{(1-R)^2}{1-2R\cos\delta+R^2},
\label{eq:etalon}
\end{equation}
where $R$ is the reflectivity at the fused silica interface, and $\delta = 4\pi l / \lambda$ is the phase shift of each subsequent round-trip reflection in the cavity, with $l$ the cavity length and $\lambda$ the wavelength of the probing light. The free spectral range of our micro-cavities is of order $\lambda$, hence we use white light to span multiple Fabry-P\'erot transmission peaks, which are fitted with a model using equation~\ref{eq:etalon} (see Appendix~A for further detail).

\begin{figure}
    \centering
    \def\svgwidth{\columnwidth}
    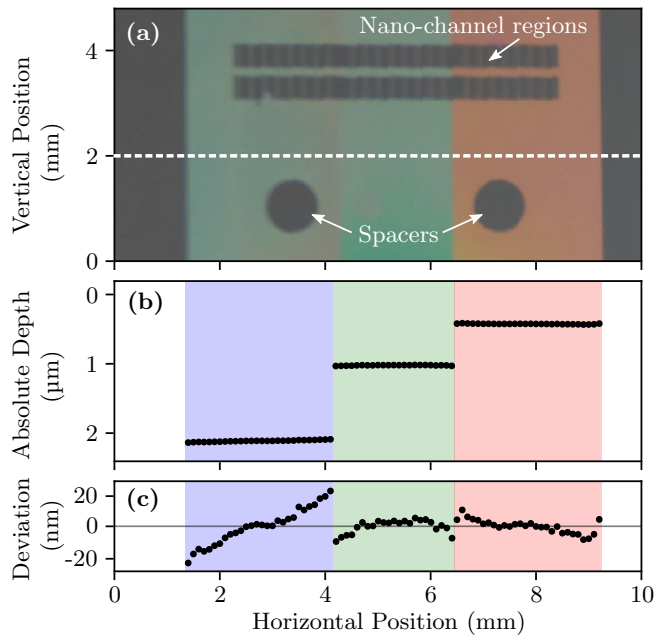
    \caption{\textbf{(a)} A close section of Fig.~\ref{fig:cell_photo}(b), showing three discrete thickness regions. The `spacers' are columns of fused silica left un-etched to prevent the cavity from bowing during evacuation. The horizontal features above are where the nano-channels are located. The dashed white line indicates the axis along which measurements in (b) were recorded. \textbf{(b)} Absolute cavity thickness (averaged over the \SI{100}{\micro\meter} focal spot size) extracted from the fitted Fabry-P\'erot transmission model as a function of lateral position across the nano-cell. The uncertainty in each depth is of order $\SI{1}{\nano\meter}$ and thus is too small to be seen. \textbf{(c)} The deviation of each datapoint from the mean value for each of the three colored sections.}
    \label{fig:WL_data}
\end{figure}

Figure~\ref{fig:WL_data}(b) shows the result of a horizontal sweep across the face of the nano-cell, with white light spectra taken at $\SI{0.1}{\milli\metre}$ intervals along the axis marked (dashed white line) in Fig.~\ref{fig:WL_data}(a). This method gives high precision measurements of cavity thickness, averaged over the \SI{100}{\micro\meter} focal spot size, at each point (with errorbars of order $\SI{1}{\nano\meter}$). From left to right, we find mean values of \SI{2070}{\nano\meter}, \SI{1020}{\nano\meter} and \SI{420}{\nano\meter} for the discrete thickness regions. The deviation from these mean values, shown in panel (c), reveals a gradual thickness change across the cell. The variations within each thickness region are of the order of $\SI{1}{\percent}$, and show that the nano-cell is slightly concave, which is likely an effect of inhomogeneous reactive ion etching.

\begin{figure}[t]
    \centering
    \def\svgwidth{\columnwidth}
    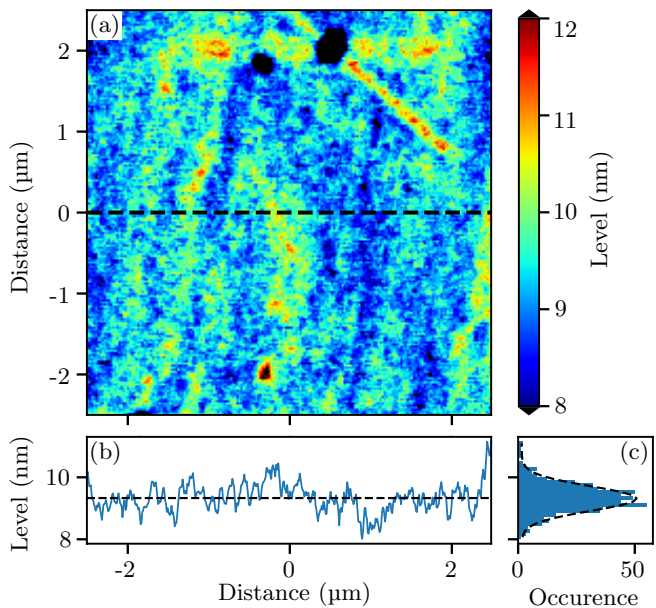
    \caption{\textbf{(a)} Image produced using AFM showing the surface roughness of a fused silica substrate etched to a depth of 500~nm. This surface was prepared in the same way as the nano-cells used in this work. This measurement was performed before the cell bonding and baking steps. The RMS roughness is 0.7~nm. If the defect at the top of the image is excluded then this becomes 0.6~nm. For comparison, an un-etched surface of the same fused silica used (not shown) gave an RMS roughness of 0.43~nm. \textbf{(b)} A horizontal cut showing the surface profile along the dashed line shown in (a). The average level is indicated (black dashed). \textbf{(c)} The distribution of points from (b) relative to the average surface level, with a Gaussian fit (black dashed).}
    \label{fig:roughness}
\end{figure}

The white light technique gives information on average thickness trends across the extent of the nano-cell. However, due to the area averaged over for each datapoint, this technique does not give information on surface roughness or defects on smaller lengthscales than the focal spot size. By using AFM on similarly prepared substrates, we find that an RMS roughness below 1~nm is routinely achievable after etching. An example is shown in Fig.\@~\ref{fig:roughness}, for a surface etched to a depth of 500~nm. Both the $\SI{5}{\micro\meter}\times \SI{5}{\micro\meter}$ region shown in (a) and the example horizontal profile shown in (b), clearly illustrate that the etched surfaces are of good optical quality. The exact roughness of each sample varies, dependent on the etch depth and conditions, and small particles or holes of order 10~nm in size may remain. However, as we have shown in this section, it is possible to control and characterize our surfaces on various lengthscales in order to understand the residual roughness, particles and gradients which may exist.

\section{Detection Schemes and Results}
\label{sec:det}

\begin{figure}
    \def\svgwidth{\columnwidth}
    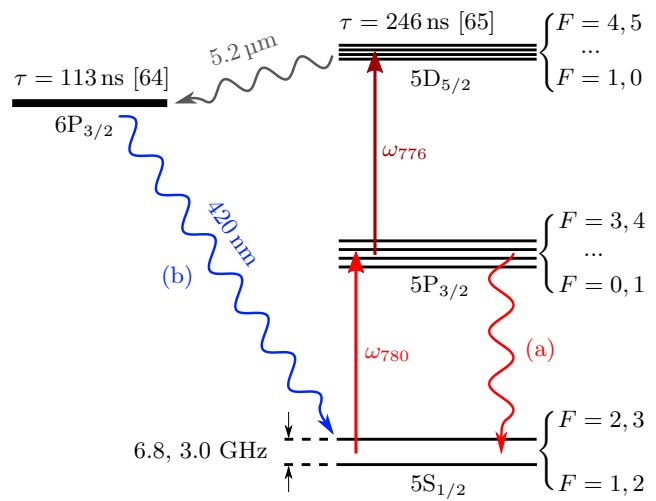
    \caption{Partial rubidium energy level diagram showing key transitions and detunings used to probe the atoms inside the nano-cell in this work. Relevant hyperfine states are indicated for $^{87}$Rb,~$^{85}$Rb (note that these are omitted for 6P$_{3/2}$ which we do not spectroscopically probe). Two detection schemes used in this work are labelled (a) and (b). Scheme (a) relies on a single excitation laser at $\SI{780}{\nano\metre}$ addressing the rubidium D2 line, driving population to the 5P$_{3/2}$ state which then decays via emission of a photon of the same frequency. Scheme (b) employs two lasers at $\SI{780}{\nano\metre}$ and $\SI{776}{\nano\metre}$ to drive atoms to the 5D$_{5/2}$ state, from which they can decay via the 6P$_{3/2}$ state and produce a photon at $\SI{420}{\nano\metre}$. The lifetimes of each of these states, $\tau$, are indicated.} 
    \label{fig:levels}
\end{figure}

The topic of detection schemes in thin cells has already received significant attention, with standard examples including transmission and selective reflection spectroscopy (see, for example, refs~\cite{Briaudeau1999,Sarkisyan2001,Sargsyan_17}), the latter having grown in popularity since the pioneering work of Woerdman and Shuurmans in probing atoms close to surfaces in 1975~\cite{Woerdman_1975}. In this section, we will begin by detailing the two detection schemes used in this work. These schemes exemplify the versatility of our vapor cell design to different input beam geometries, and are used in this study to spectroscopically probe rubidium atoms within our nano-cells.

The energy levels of rubidium relevant to the detection methods used in this work are shown in Fig.\@~\ref{fig:levels}. In the most straightforward scheme, a single tunable laser at $\lambda = \SI{780}{\nano\metre}$ is used to excite Rb atoms from their ground state 5S$_{1/2}$ to the excited state 5P$_{3/2}$, i.e.\@~on the D2 line. Extinction spectroscopy in transmission on this (or the D1 line, not shown in Fig.\@~\ref{fig:levels}) has been the most widely adopted detection technique for studying thermal vapors due to the simplicity of implementation. However, when the optical depth of the system is decreased, as in the case of a nano-thickness vapor, it becomes more challenging to detect the extinction signal due to reduced signal-to-noise~\cite{Taylor-NL19}. Our geometry allows us to instead perform fluorescence measurements efficiently, employing a high-NA objective lens to collect atomic fluorescence through the thin cover glass. 

Two fluorescence-based detection schemes are used in this work. The single-photon scheme labelled (a) in Fig.~\ref{fig:levels} allows for detection of fluorescence activity with the excitation laser light spatially filtered from the atomic fluorescence, for example in a total internal reflection fluorescence (TIRF) geometry (see Fig.~\ref{fig:TIRF}). Alternatively, a two-photon excitation scheme can be used. A $\SI{776}{\nano\metre}$ laser is added to drive population to the 5D$_{5/2}$ state, from which there is approximately~$\SI{7.5}{\percent}$ probability that an atom will decay via the 6P$_{3/2}$ state and produce a $\SI{420}{\nano\metre}$ photon \cite{Sheng2013} (labelled (b) in Fig.~\ref{fig:levels}). This light can be spectrally separated from scattered excitation laser light, providing an even higher signal-to-background ratio than in the single-photon scheme.

Below we discuss the two detection schemes in more detail, and employ them to record spectroscopic data from atoms confined to the nano-cell. In all cases the frequency axis was calibrated by a reference Rb cell in a methodology similar to that performed in reference~\cite{Siddons2008}, where the zero of detuning is chosen to be the weighted line-center of the rubidium D2 transition.

\subsection{Total Internal Reflection Fluorescence (TIRF)} \label{sec:TIRF}

\begin{figure}
    \centering
    \def\svgwidth{\linewidth}
    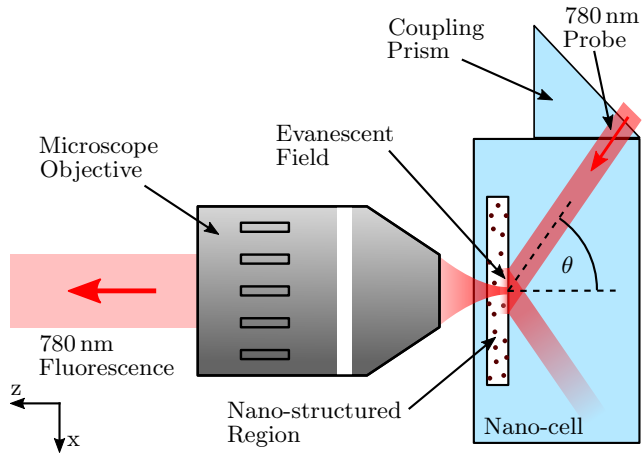
    \caption{A top-down schematic diagram of the TIRF setup. A coupling prism is contacted to the cell with index-matching oil, and this is used to deliver a $\SI{780}{\nano\metre}$ probe beam into the nano-cell at an angle $\theta$ to the normal. When incident on the back interface between the nano-region and the glass, the beam totally internally reflects and an evanescent field is formed inside the nano-region. This field excites the atomic vapor. Fluorescence is collected through the front face of the cell via a $f=\SI{2}{\milli\meter}$ focal length, $\textrm{NA} = 0.7$ plan apochromat objective lens, onto a single-photon avalanche diode. Note that the coupling prism used is the same for all datasets, and simply improves ease of alignment at more extreme angles.}
    \label{fig:TIRF}

\end{figure}

\begin{figure}[ht]
    \centering
    \def\svgwidth{\columnwidth}
    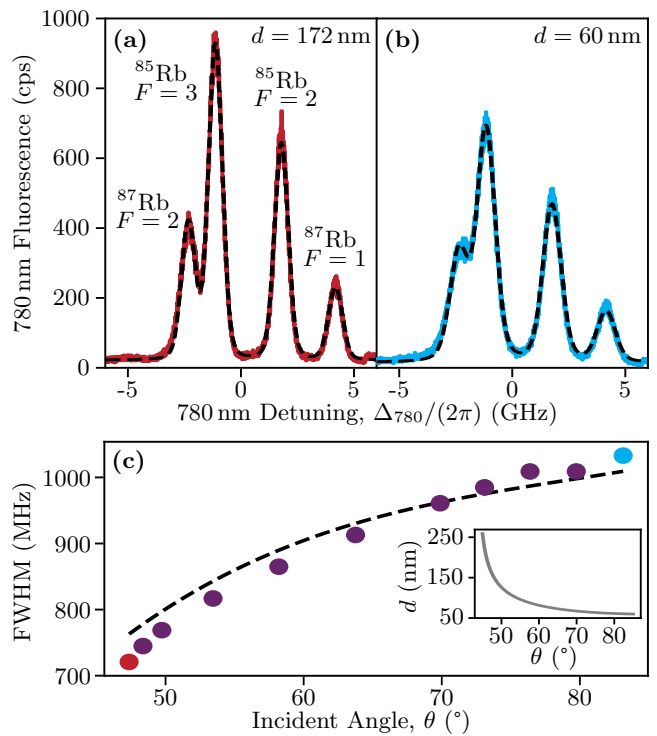
    \caption{\textbf{(a)} A TIRF spectrum (red) recorded with incident angle $\theta = \SI{47}{\degree}$ onto a $\SI{2}{\micro\metre}$ thick vapor layer, with fitted Voigt model (black dashes). The spectrum has narrow line-widths due to the long evanescent field depth ($d = \SI{172}{\nano\meter}$) resulting in less transit-time broadening. The hyperfine ground states corresponding to each resonance have been labelled for clarity. \textbf{(b)} TIRF spectrum (blue) as in (a), but with $\theta = \SI{83}{\degree}$ resulting in a shorter evanescent field ($d = \SI{60}{\nano\meter}$) and thus greater transit-time broadening. Fitted model shown as black dashed line. \textbf{(c)} Measured FWHM extracted from the Voigt model for TIRF spectra as a function of the $\theta$ (colored dots). Horizontal errorbars cannot be seen but are of order 1$\SI{}{\degree}$. The extreme data points are color matched to the spectra in (a) and (b). Also plotted is the result of a Monte Carlo simulation of thermal atoms travelling within the detection volume (black dashed line). Inset shows the evanescent decay length as a function of $\theta$ (see Eq.~\ref{eq:d}). 
    }
    \label{fig:angle}
\end{figure}

A schematic of the TIRF setup is shown in Fig.~\ref{fig:TIRF}. Total internal reflection occurs at the glass-vapor interface, and a single-color evanescent light field is created that can drive atomic excitation and induce fluorescence. Although strictly 3-dimensional, we consider only the variation of the laser field along the axial distance $z$ from the interface between the fused silica wall and the vapor. Hence, the evanescent field has the general form $I(z)=I_0 e^{-z/d}$, where $I_0$ is the intensity at the boundary immediately outside of the medium in which the total internal reflection is occurring, $z$ is the perpendicular distance from the fused silica wall into the vapor, and $d$ is the evanescent field $1/e$ decay length. This is given by~\cite{Axelrod1984}:
\begin{equation}
    d = \frac{\lambda}{4\pi} \Big( n_1^2\mathrm{sin}^2\theta-n_2^2 \Big) ^{-\sfrac{1}{2}},
\label{eq:d}
\end{equation}
with $n_1$ the refractive index of fused silica ($\approx\SI{1.45}{}$), $n_2$ the refractive index of the vapor (nominally taken as $\SI{1}{}$), $\theta$ the incident angle, and $\lambda$ the vacuum wavelength of the light field (we probe the D2 line at $\lambda$ = $\SI{780}{\nano\metre}$). Typically we have $d < \lambda/2\pi$, and thus expect frustrated total internal reflection to have little effect in our nano-cells with internal dimensions $\sim \lambda$.

To demonstrate the TIRF methodology, experiments were performed to study the transit-time broadening induced by varying evanescent field decay lengths. The nano-cell was heated to $\SI{100}{\degree}$C, and fluorescence counts were collected for 10~s with a probe beam power of~\SI{10}{\micro\watt}. The resulting fluorescence spectra shown in Fig.\@~\ref{fig:angle} contain transitions from each of the two hyperfine 5S$_{1/2}$ ground states, for each of the two naturally occurring rubidium isotopes, to the 5P$_{3/2}$ state (see Fig.\@~\ref{fig:levels})~\cite{Siddons2008}. The hyperfine structure of the 5P$_{3/2}$ state, with features $\sim 100$~MHz, is not resolved due to Doppler broadening. 

To analyze the observed spectra, we consider a line-shape model made by the summing of four Voigt profiles, each centered on one of the four hyperfine ground states observed in the fluorescence spectrum. The full width at half maximum (FWHM) of one Voigt profile is plotted as a function of $\theta$ (colored dots, Fig.~\ref{fig:angle}(c)). The trend reveals that the spectra become broader as the incident beam angle is increased. We understand this to be a transit-time broadening caused by the reduced characteristic decay length, $d$, of the evanescent light field that probes the atoms (see Fig.~\ref{fig:angle}(c) inset).

To examine the observed line-width behavior further, we performed a Monte Carlo simulation. The simulation initializes 3000 individual atoms with a random velocity vector $\vec{v}=v_x\hat{x}+v_y\hat{y}+v_z\hat{z}$ (see Fig.~\ref{fig:TIRF} for axes definition) such that the distribution of velocity vectors approximates the Maxwell-Boltzmann distribution as expected for an ideal thermal vapor. The spectrum of the ensemble is made up of Lorentzian line-shapes contributed by individual atoms, which are modified in two ways. First, the line-center is shifted by the Doppler effect $\vec{k}\cdot\vec{v}=kv_z$ as the atom moves with respect to the laboratory frame. Second, the width of the Lorentzian is modified by the short transit time $t_{\mathrm{trans}}=d/v_z$ across the short decay length $d$ of the evanescent field along $z$. To put this timescale in perspective, we note that at a typical temperature of \SI{100}{\degree}C a thermal atom has a mean speed of 330~m\,s$^{-1}$, corresponding to $t_\textrm{trans} = 0.3$~ns across a distance $d = 100$~nm. We define an effective line-width $\Gamma_{\mathrm{eff}}$ per atom using:
\begin{equation}
    \Gamma_{\mathrm{eff}}^2 = \Gamma_{\mathrm{nat}}^2 + \bigg( \frac{\alpha}{t_{\mathrm{trans}}} \bigg) ^2,
\end{equation}
where $\Gamma_{\mathrm{nat}}$ is the natural line-width ($2\pi\times6$~MHz~\cite{Volz_1996}), and $\alpha$ is a unitless factor that attenuates the transit-time broadening. We do not consider atomic interactions or collisions, as typically we have atomic densities of order $10^{12}$~cm$^{-3}$ and do not have buffer gas within our cells. The 3000 atom simulation was repeated ten times at a number of discrete values of incident beam angle $\theta$ to return a mean and standard error for each. A fit to the simulated data is plotted as the black dashed line in Fig.~\ref{fig:angle}(c). We find that with $\alpha = 0.5$ the simulation captures the experimentally observed trend, however clearly some systematic error exists. For comparison, for transit-time broadening of atoms traversing a Gaussian laser beam the comparable factor is $\alpha \approx 0.2$~\cite{LaserSpec}. More complex modelling is beyond the scope of this work, however, we speculate that the differences between observation and theory come about due to the evanescent field gradient experienced by the atoms, as well as the reduced probability of a fast moving atom being excited by the laser field. Neither of these effects were included in our model. The latter means that atoms with a low transit time are less likely to contribute to the fluorescence signal, reducing the overall broadening effect.

\subsection{Two-Photon Fluorescence Microscopy (TPFM)}

As illustrated in the previous section, the technique of single-photon TIRF gives a high signal-to-background ratio for spectroscopy. However, it relies on detection of fluorescence photons at the same frequency as the input laser photons, so although it is theoretically a dark-field method, it is in practice not free of background scattering. This poses a problem in particular for cells with more complex structures such as one-dimensional nano-channels (see, for example, Fig.~\ref{fig:cell_photo}(c)), where laser light is efficiently scattered by the edges. In this section, we detail a two-photon excitation scheme which allows fluorescence detection at 420~nm. The scheme is depicted in Fig.~\ref{fig:levels}, with the fluorescence pathway used for detection labelled~(b). This scheme allows for straightforward filtering of any excitation photons scattered by the substrate, using commercially available bandpass filters.

\begin{figure}[t!]
    \centering
    \def\svgwidth{\columnwidth}
    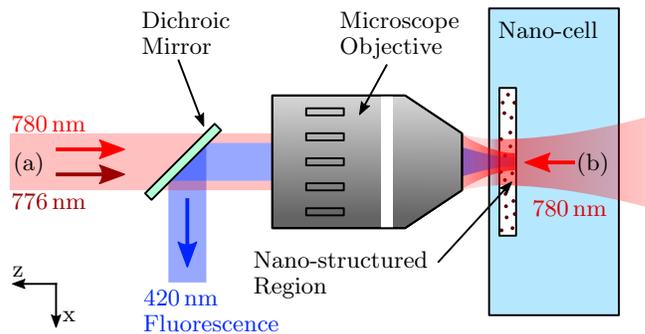
    \caption{Top-down schematic of the fundamental components in the TPFM setup. Probe lasers at \SI{780}{\nano\metre} and \SI{776}{\nano\metre} are delivered to the nano-regions of the vapor cell, with the \SI{780}{\nano\metre} beam co-propagating (a) or counter-propagating (b). Beams along pathway (a) are focused into the nano-cell by a $100\times$ plan apochromat microscope objective, with $\SI{2}{\milli\metre}$ focal length, $\SI{6}{\milli\metre}$ working distance, and $\textrm{NA} = 0.7$. Along pathway (b) the beam is weakly focused through the back face of the cell using an $f=100$~mm lens. Atoms inside the nano-structures produce $\SI{420}{\nano\metre}$ fluorescence (see excitation scheme in Fig.~\ref{fig:levels}) which is collected by the microscope objective, and reflected by a dichroic mirror onto a photon-counting photo-multiplier tube (PMT) for detection.}
    \label{fig:tpfm}
\end{figure}

\begin{figure}[t!]
    \centering
\def\svgwidth{\columnwidth}
    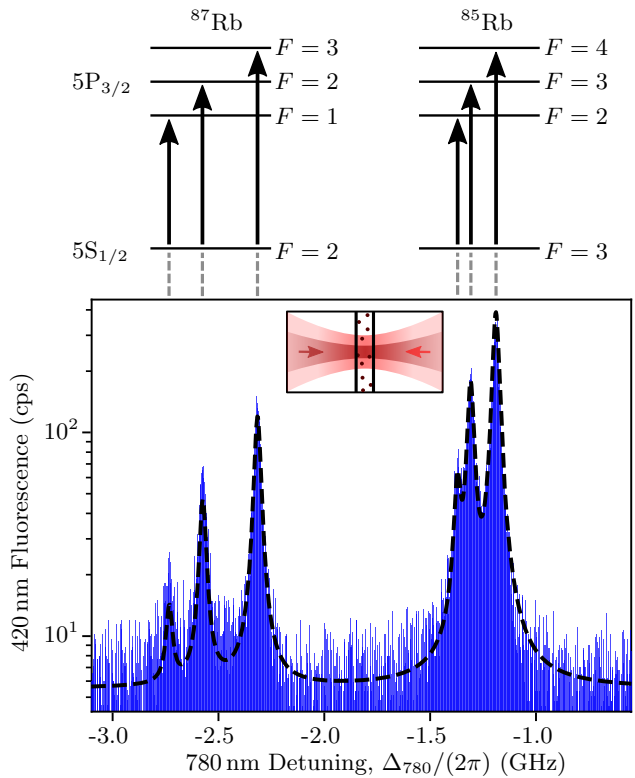
    \caption{\SI{420}{\nano\meter} fluorescence (blue histogram) recorded from a region of the nano-cell with \SI{1}{\micro\meter} depth (i.e.\@~confinement along $z$). A counter-propagating two-photon excitation geometry is used (see inset diagram illustrating atomic layer and input beams, as well as Fig.\@~\ref{fig:tpfm}). Note that this spectrum corresponds to the left half of that shown in Fig.\@~\ref{fig:angle}(a). An empirical fit comprising six Voigt profiles (black dashed) was performed, with the locations of the resonances constrained to the relevant hyperfine transition frequencies, their magnitudes constrained to the isotopic abundance-weighted transition strengths, and their widths set to be equal (but allowed to vary as a fit parameter). This yields a line-width of $(32\pm1)$~MHz. Partial energy level diagrams are shown above for reference. For this dataset, powers of $\SI{350}{\micro\watt}$ at 780~nm (focussed with the $f=100$~mm lens) and \SI{20}{\micro\watt} at 776~nm (focussed with the $f=2$~mm lens) were used, with a cell temperature of \SI{60}{\degreeCelsius} and an integration time of 12~h.}
    \label{fig:counterprop}
\end{figure}

Our nano-cell design allows for the use of high-NA optics, which allows us to optically resolve nano-scale structures, to address and collect fluorescence from atoms within individual structures, and to achieve extremely high Rabi frequencies even on weak transitions. In this section we employ a $\textrm{NA} = 0.7$ plan apochromat microscope objective to address atoms within volumes $\sim\lambda^3$ confined in nano-scale structures, where spatial confinement is typically $\leq\SI{1}{\micro\meter}$ in one or two dimensions. Figure~\ref{fig:tpfm} shows the experimental setup, where the two excitation lasers are delivered to the nano-regions in co- or counter-propagating arrangements, and atomic fluorescence photons are collected via the microscope objective.

Our high-NA approach leads to small beam sizes and extremely high Rabi frequencies at modest laser power. For example, a 776~nm laser power of \SI{20}{\micro\watt} and a spot size of 1 micron gives a Rabi frequency of 2.9~GHz on the 5P$_{3/2} \rightarrow \textrm{5D}_{5/2}$ transition (using the dipole matrix element measured for this transition in~\cite{Whiting_2016}). This means that atoms in the 5P$_{3/2}$ state are promoted to the 5D$_{5/2}$ state at a rate much faster than the Doppler width. Below we will discuss the effect of this on resonance line-shapes observed in both co- and counter-propagating regimes.

In the experiment, we scan the first step 780~nm laser through the D2 resonance while employing excited-state polarization spectroscopy~\cite{Carr_12} to frequency stabilize a second step laser at 776~nm to the 5P$_{3/2} \rightarrow \textrm{5D}_{5/2}$ transition. Subsequently we detect 420~nm fluorescence photons from the $\textrm{6P}_{3/2} \rightarrow \textrm{5S}_{1/2}$ decay (see Fig.~\ref{fig:levels}).

Firstly, we report on the results of implementing a TPFM scheme in which the two excitation beams are counter-propagating in the nano-cell (see (b) in Fig.\@~\ref{fig:tpfm}). As can be seen in Fig.~\ref{fig:counterprop}, this scheme gives well resolved sub-Doppler fluorescence resonances corresponding to the transitions from the hyperfine 5S$_{1/2}$ ground states to the allowed 5P$_{3/2}$ hyperfine states. In this scheme, a large 780~nm excitation volume allows atoms to be excited to the 5P$_{3/2}$ hyperfine states. Some of these excited atoms will subsequently enter the tightly focussed 776~nm spot, and will then sequentially excite to the 5D$_{5/2}$ state due to the high Rabi frequency as noted above. Hence, the splittings and line-strengths observed correspond closely to those for the 5P$_{3/2}$ hyperfine levels. A Voigt fit yields a  FWHM of ($32\pm1$)\,MHz, consistent with transit-time broadening with \SI{1}{\micro\meter} confinement. In contrast to single laser pump-probe schemes, crossover resonances are also absent, affording better resolution of the allowed hyperfine transitions. Note that nano-cells are the ideal platform for the study of the interaction of atoms and high intensity laser fields, as the extent of the confined atoms can be smaller than the Rayleigh range, thus removing the necessity to average over different intensity regimes (which would be the case in a standard millimeter-sized vapor cell). Our bespoke nano-cell design allows high-NA objectives to be placed close to the atomic layer to facilitate such studies. 

\begin{figure}[t!]
    \centering
\def\svgwidth{\columnwidth}
    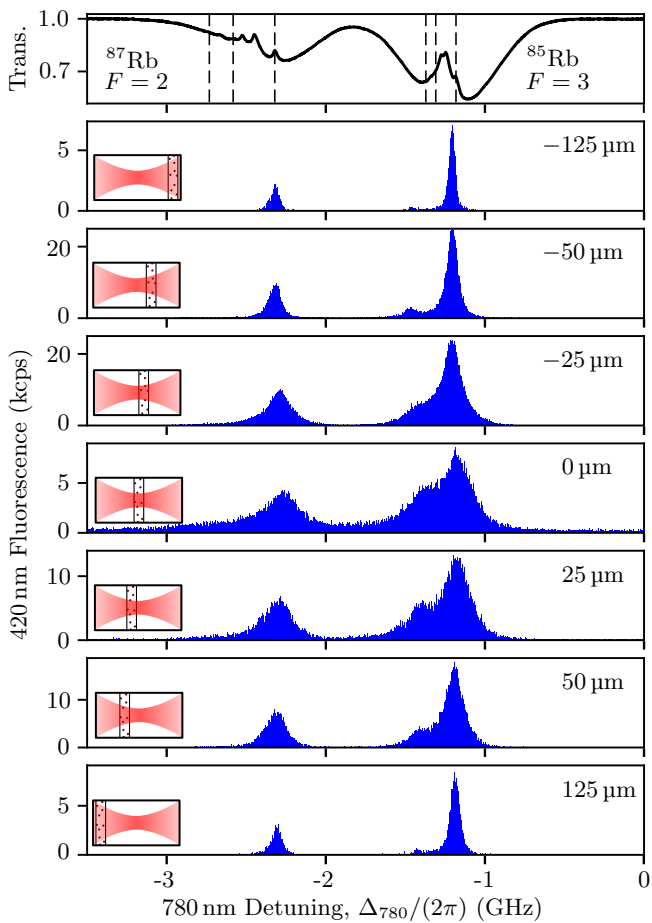
    \caption{\textbf{Top:} Reference spectrum produced using a single-color pump-probe hyperfine spectroscopy setup~\cite{Smith_2004} in a 75~mm cell. Each Doppler-broadened profile exhibits six sub-Doppler features (some too weak to be seen), with the three indicated corresponding to the hyperfine 5P$_{3/2}$ states (see Fig.\@~\ref{fig:counterprop}), and the other three to crossover resonances between these. \textbf{Below:} TPFM spectra recorded with 780~nm and 776~nm beams co-propagating (both along path~(a) in Fig.~\ref{fig:tpfm}) incident on a region of the nano-cell with \SI{1}{\micro\meter} depth in the direction of the excitation beams. The nano-cell was translated with respect to the focal point of the beams (labelled \SI{0}{\micro\meter}), such that the atoms experience different intensity environments at different $z$ positions (labelled). Inset diagrams illustrate this translation of the atomic layer with respect to the focus. High intensities at the focus give rise to power broadening, which in combination with the changing collection efficiency and illumination area, also results in the changing count rates between spectra. Features in the TPFM spectra relate to the hyperfine 5P$_{3/2}$ states. Excitation beams had powers of $\SI{7}{\micro\watt}$ at 780~nm and \SI{20}{\micro\watt} at 776~nm, with a Rayleigh range of approximately $\SI{4}{\micro\meter}$. Each dataset had an integration time of 1--10~mins at a cell temperature of \SI{65}{\degreeCelsius}.
    }
    \label{fig:defocussing}
\end{figure}

Spectra were also obtained using the TPFM method with a co-propagating geometry (see (a) in Fig.~\ref{fig:tpfm}), and a selection of these are shown in Fig.~\ref{fig:defocussing}. Here the nano-cell was displaced along the optical axis of the microscope objective, such that atoms sampled the laser intensity profile locally. The spectral features in the datasets arise due to the hyperfine structure of the 5P$_{3/2}$ state, though not all hyperfine states are fully observed or resolved. We no longer observe the same activity ratios between hyperfine levels as in counter-propagating case (Fig.\@~\ref{fig:counterprop}). We attribute this to the fact that, in this scheme, both lasers are tightly focussed, leading to a small excitation volume. This means atoms must be excited to the 5D$_{5/2}$ state within the timescale of transit across the \SI{1}{\micro\meter} spot (approximately 4~ns). In contrast to the sequential excitation in the counter-propagating case, signal is now only observed where the 780~nm laser is at the correct frequency such that it can, in combination with the frequency-stabilized 776~nm laser, cause atoms to be excited to the 5D$_{5/2}$ state. At the higher intensities studied we do observe a secondary resonance to the left of the main $^{85}$Rb peak, but in this case the resonances are strongly power broadened and Stark shifted.

The data in Fig.\@~\ref{fig:defocussing} clearly illustrates that the spectroscopy method used is very sensitive to changes in position of the thin atomic medium. The observed modification of the spectra can be attributed to two main effects. Firstly, power broadening causes broader spectral features closer to the focus ($z=\SI{0}{\micro\meter}$) due to the higher intensities experienced by the atoms. Secondly, moving away from the focus alters the amount of fluorescence generated by the atoms (via changing intensity and illuminated area) and also collected by the objective lens, causing the changes in count rates observed. Transit time broadening due to the small spot size at the focus will also be a less significant contributing factor. The spectra which do not strongly exhibit the effects of power broadening (for example those at $z=\pm\,\SI{125}{\micro\meter}$) also illustrate the benefit of our experimental scheme: narrow sub-Doppler fluorescence resonances are obtained in a co-propagating geometry as well as the counter-propagating one seen in Fig.~\ref{fig:counterprop}. This allows for a compact setup whereby a single objective delivers both excitation beams to a tightly focused spot in the atomic medium, as well as collecting the atomic fluorescence. In fact, a Voigt fit to the uppermost spectrum ($z=\SI{125}{\micro\meter}$) yields a FWHM of ($57.5\pm0.6$)\,MHz, the same order of magnitude as that achieved in the counter-propagating case.

\begin{figure}[t!]
    \centering
\def\svgwidth{\columnwidth}
    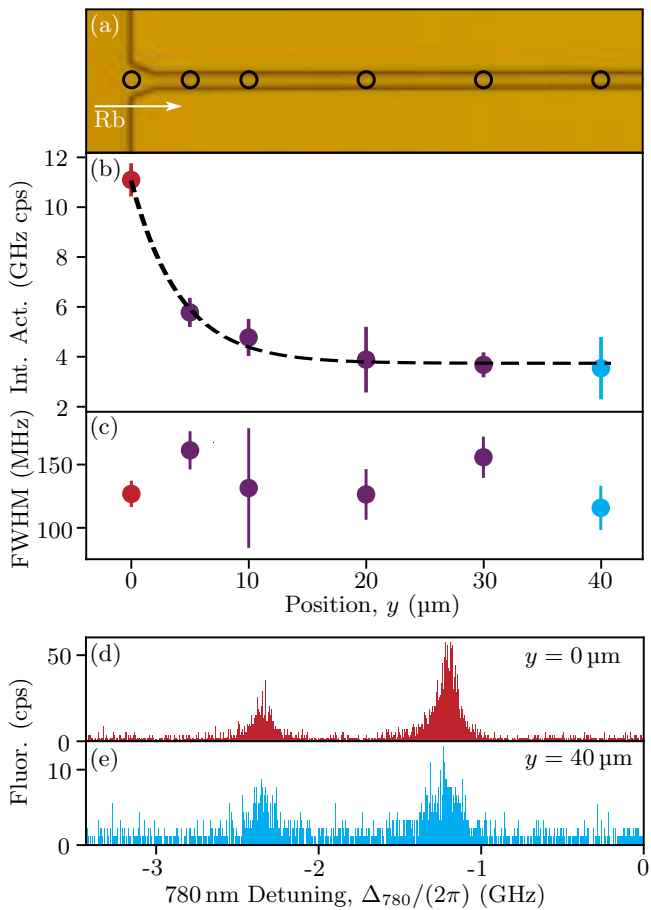
    \caption{\textbf{(a)} Bright-field image of a single channel, with a width and depth of $\SI{1}{\micro\meter}$. In this image the source of rubidium atoms (reservoir) is from the left, as indicated. TPFM spectra were recorded at various locations along this channel (indicated). All spectra were empirically fitted with Voigt profiles, and from these fits the \textbf{(b)}~integrated count rates and \textbf{(c)}~FWHMs are plotted as a function of position. An empirical exponential decay is fitted to the activity data (black dashed), which yields a characteristic decay length of ($4\pm1$)\,$\SI{}{\micro\meter}$. Errorbars for both the activity and FWHM values were estimated using the functional approach~\cite{HughesHase}. \textbf{(d)} and \textbf{(e)} show two example spectra, color matched to the extreme points studied. This dataset was taken with powers of $\SI{50}{\nano\watt}$ at 780~nm and 776~nm, integration times of 0.5--9~hours, and a cell temperature of \SI{60}{\degreeCelsius})
    }
    \label{fig:channel_data}
\end{figure}

The TPFM method in the co-propagating geometry was also used to probe atomic vapor confined along a channel with a width and depth of $\SI{1}{\micro\metre}$ (shown in Fig.\@~\ref{fig:channel_data}(a)). As displayed in Fig.\@~\ref{fig:channel_data}(b), we find that a density gradient exists in the vapor as the fluorescence intensity reduces along the channel (with all other experimental parameters held constant between measurement sites). Given that this particular cell had been in operation for 1 year, it would be reasonably expected that an effective steady state had been reached in terms of atomic vapor distribution. However, as is shown by the decaying integrated count rate, this is not the case. Instead we have found that atomic vapor density is reproducibly not uniform along channels with confinement on lengthscales $\leq \SI{1}{\micro\meter}$. The vapor pressure is perturbed simply by the presence of tight confinement, a result which provides new and important insight into the diffusion properties of atoms inside nano-scale structures.

For the experiments reported in Fig.\@~\ref{fig:channel_data}, the excitation beam powers used were a compromise between signal strength and power broadening. 
Hence by fitting Voigt profiles to the spectra we obtain an average FWHM of 140~MHz and, as illustrated in Fig.\@~\ref{fig:channel_data}(c), show that the FWHM does not vary significantly between sites in this tight confinement regime. The observed power broadening alludes to the tight focus of the beam which delivers a high intensity even at the low powers used. In Fig.\@~\ref{fig:channel_data}(d) and (e), we present spectra corresponding to the closest and farthest measurement points, which have FWHMs of ($130\pm10$)\,MHz and ($120\pm20$)\,MHz, respectively.

The TPFM method offers promise for the detection and study of low numbers of atoms confined to nano-scale structures. From the count rates observed in the spectrum shown in Fig.~\ref{fig:channel_data}(e), we have extracted an estimate of the number of atoms excited to the 5D$_{5/2}$ state. Accounting for the efficiency and throughput of our detection setup, as well as the lifetime of the 5D$_{5/2}$ state, we estimate that the on resonance count rate observed corresponds to a mean value of just 0.01 atoms excited to the 5D$_{5/2}$ state within the detection volume. If instead we extract the atomic number density from the measured cell temperature and the rubidium vapor pressure curve, as is standard in experiments with bulk vapor cells~\cite{Alcock1984}, we find a mean ground state atom number of 1.5 within the same volume. Whilst we do not believe the bulk number density relationships to exactly hold in a thin cell (a consequence of tight confinement which our results in Fig.~\ref{fig:channel_data} begin to shed light on), this calculation provides an estimate for comparison. The extracted value of 1.5 ground state atoms is consistent with the experimental value of 0.01 atoms in the 5D$_{5/2}$ state, as we would expect that only a fraction of atoms are excited to this state. Further modelling is, however, beyond the scope of this work.

Our data have also shown that it is possible to spectrally filter fluorescence signals from the TPFM method in such a way that the noise observed far off resonance, such as that in the spectra shown in Fig.~\ref{fig:counterprop} and Figs.~\ref{fig:channel_data}(d) and (e), is consistent with the dark count rate measured for the photon-counting PMT used in this work. This dark count rate is approximately $\leq5$~cps, and the input laser powers used in this work lie in the range $\SI{}{\micro\watt}$--$\SI{}{\nano\watt}$. As such this method is highly sensitive to low atom numbers (exemplified by the estimated atom number and moderate cell temperatures used in this section), whilst also allowing for long integration times limited only by dark noise. Thus, given our demonstrated experimental versatility and considerable control over atom numbers through cell temperature and confinement geometries, work towards interrogating low numbers of atoms is a promising avenue of further study.

\section{Conclusion}
We have designed and fabricated a nano-structured alkali-metal vapor cell that offers flexible optical access and high-NA imaging. Our glue-free design has proven to be durable and reliable, avoiding out-gassing and degradation of the internal atmosphere over time. The formation of nano-scale structures inside our cell, via direct write laser lithography and reactive ion etching, is flexible and highly customizable. 

We have illustrated the versatility of optical access afforded by our nano-cell design through the methods of TIRF and TPFM. 
The TIRF method allows for fine control of the atomic excitation region on the nanometer-scale, enabling our study of the transit-time broadening induced on this lengthscale. Meanwhile, TPFM is a high signal-to-noise dark-field measurement, and is used in this work to illustrate the potential of nano-cells for sensing applications by studying the spectral responses of the vapor layer. Here our high-NA focusing also enables strong driving of even weaker transitions, and it is this effect, rather than Doppler broadening, that dominates the observed line-shapes. Finally, using the TPFM method, we have spectroscopically probed atoms confined by $\SI{1}{\micro\meter}$ in two dimensions. Through this method, we have demonstrated that tight confinement regimes modify the local atomic vapor density, paving the way for development towards room temperature atom-based devices with close control of atom number at the level of single atoms.

The highly localized control of atom number available by patterning confinement geometries that we have demonstrated opens doors towards the possibility of few- and single-atom applications. For example the local vapor density could be controlled in such a way as to produce a single atom source, for example by patterning nano-channel geometries to, on average, only allow a single atom from the thermal ensemble to pass through. This could open doors to heralded single atom generation. Engineering local atomic density towards having only a single atom on average within an interaction region would then hold promise for work towards single photon sources, for which our platform would provide a scalable and robust architecture.

The nanoscopic character of our vapor cell and the omnipresence of its walls readily provide sub-Doppler spectral resolution, giving access to individual hyperfine transitions of rubidium. Other technical advantages of our cells include the integrated ITO heating system and the high excitation and collection efficiency of our high-NA approach, which allow for operating the chip at relatively low temperatures compatible with biological sensing. The combination of these features with high-NA lateral resolution and the ultra-thin extent of the atoms makes our nano-cell a powerful spatially-selective sensing tool. A special example of current interest in sensing is magnetometry with nano-scale resolution~\cite{Maze2008}, and indeed sensitive magnetometry has been recently demonstrated using nano-cells~\cite{Klinger2019}. Our chip and measurement platform allows for improved spatial selectivity and high lateral resolution, especially for probing magnetic fields close to surfaces. The thin front panel and compactness of our design are key benefits in this regard, especially with regards to the ITO heater which removes the problem of bulky cell ovens experienced with previously reported vapor cells. That we can pattern arbitrary nano-structured geometries, rather than the standard wedge-shape, also gives considerably greater versatility for extremely spatially-selective measurement (e.g.\@ spatially-selective field imaging in one dimension using a thin atomic layer, or in two dimensions using a patterned nano-channel. The bespoke nature of our novel fabrication process also makes it possible, for example, to deposit material layers during cell manufacture which ultimately reside inside the finished cell. This could allow measurement of magnetic fields extremely close to material surfaces, enabled by the proximity of the atomic nano-layer. 

The architecture of our nano-cell makes it well adapted to the application of microscopy and spectroscopy techniques that are routine in nano-optics. As well as TIRF and TPFM, further potential examples include stimulated emission depletion (STED) fluorescence microscopy~\cite{Hell1994}; fluorescence correlation spectroscopy (FCS)~\cite{Elson1974}; and interferometric scattering (iSCAT) - a scheme which has been used previously to detect single nanoparticles and even charge transport via Rayleigh scattering~\cite{Taylor-NL19, Delor-19}. These detection schemes, along with the versatility of the cell design and manufacturing process, offer access to a multitude of investigative routes. Further potential directions of study include the use of light-induced atomic desorption (LIAD) to locally modify the atomic vapor density, depositing micro-electrodes or micro-patterning structures such as waveguides or microring resonators within the nano-cells, or even adding dipole trapping to the setup~\cite{Sandoghdar1996}. Maturing the platform with integration of optical components to an on-chip design offers a promising route towards highly-scalable atom-based quantum technologies.
\\

\section*{Acknowledgements}
The authors are grateful to Dr.~Sofia de~Carvalho Ribeiro for editorial discussions. We also thank Benjamin Gmeiner and Pierre T\"urschmann for their contributions to the early phase of this work. We gratefully acknowledge support by the Max Planck Society and EPSRC grant EP/R002061/1. Correspondence should be sent to vahid.sandoghdar@mpl.mpg.de or c.s.adams@durham.ac.uk.

T.F.C.\@ and W.J.H.\@ contributed equally to this work. 

The data presented in this paper are available from Durham Research Online, doi:10.15128/r1g732d901w

\bibliography{refs}

\end{document}